\documentclass[12pt,preprint]{aastex}
\usepackage{lscape}
\def\kms{{\rm km\,s^{-1}}}

\def\masyr{{{{\rm mas}}\,{{\rm yr}}^{-1}}}
\def\usno{{\rm USNO}}

\def\lim{{\rm lim}}

\def\bias{{\rm bias}}
\def\tot{{\rm tot}}
\def\cor{{\rm cor}}
\def\cov{{\rm cov}}
\def\ext{{\rm ext}}
\def\inter{{\rm int}}
\def\u{{\rm U}}
\def\s{{\rm S}}
\def\l{{\langle}}
\def\r{{\rangle}}
\def\smu{{\sigma_\mu}}

\begin{document}

\title{Proper Motion Catalog From SDSS $\cap$ USNO-B}

\author{Andrew Gould and Juna A.\ Kollmeier}
\affil{Department of Astronomy, The Ohio State University,
140 W.\ 18th Ave., Columbus, OH 43210}
\authoremail
{gould,jak@astronomy.ohio-state.edu}

\singlespace

\begin{abstract}

We present a catalog of 345,000 stars with proper motions 
$\mu\geq 20\,\masyr$ and magnitudes $r'\leq 20$,\
drawn from the intersection of the SDSS Data Release One and USNO-B. 
We first use SDSS quasars to correct the proper motions
in each of the two source catalogs, finding that these corrections
for systematics are comparable to the statistical errors.  The
combined proper motions have errors of $\sigma_\mu=4.0\,\masyr$,
significantly better than any previous catalog at these magnitudes.
While cross-correlating SDSS and USNO-B removes the vast majority
of the very large number of spurious proper-motion stars in each,
some contamination remains, especially at faint magnitudes, $r'\ga 19.5$,
and high proper motions.  We present a diagrammatic method to estimate
the level of contamination in selected subsets of the catalog, which
should be useful for choosing appropriate selection criteria for
various applications.

\end{abstract}
\keywords{astrometry -- catalogs -- stars: fundamental parameters
-- stars: subdwarfs -- late-type -- white dwarfs}
 
\section{Introduction
\label{sec:intro}}

Combining stellar proper motions with photometry has long been an
efficient method to select otherwise elusive populations of stars.
Proper motion surveys can be broadly divided into two classes: those
that cover a large fraction of the sky and those that cover small
areas but have faint limiting magnitudes.

All-sky surveys generally trade magnitude limits for proper motion
precision or threshold.  For example, the combined Hipparcos
\citep{esa97} and Tycho 2 \citep{hog00} catalogs provide proper
motions accurate to $\smu \la 1\,\masyr$ for essentially all
stars with $V \la 12$. The nearly completed USNO CCD Astrograph
Catalogs (UCAC1,2) \citep{zacharias00, zacharias01} have $\smu$
between $1\,\masyr$ and $5\,\masyr$ depending on magnitude for $V\la 16$.
The New Luyten Two Tenths (NLTT) Catalog
\citep{luyten7980,luytenhughes80} is reasonably complete to $V\sim19$
over most of the sky, but at the price of being restricted to high
proper-motion stars ($\mu>180\,\masyr$).  Moreover, at $\smu\sim 20\,\masyr$
\citep{salim03}, its accuracy is substantially worse than catalogs at
brighter magnitudes.  However, the revised NLTT (rNLTT) improves this
precision to $\smu=5.5\,\masyr$ for 44\% of the sky \citep{gould03a,
salim03}.  Most recently, \citet{monet03} have analyzed 4 decades of
photographic plates to produce the monumental USNO-B, an all-sky
proper-motion catalog with typical $\smu \sim 7\,\masyr$ for all stars
(no proper-motion threshold) down to $V\sim21$.  However, while this
catalog is $90\%$ complete for high proper motion stars ($\mu\ga
180\,\masyr$), about $99\%$ of its entries in this regime are spurious
\citep{gould03e}, so it cannot be directly accessed as a source of high
proper motion stars.

Due to the relatively bright magnitude limits of these all sky surveys
(other than USNO-B), other groups have been motivated to do
smaller-area, deeper studies, focusing instead on a particular 
proper-motion or magnitude limit as opposed to all-sky coverage. The
SuperCOSMOS catalog \citep{hambly01a, hambly01b}, covering a 5000 square
degree patch of the southern Galactic cap, obtains proper motion
measurements with $\smu \sim 10\,\masyr$ at $R\sim18$, and $\smu
\sim 50\,\masyr$ at $R\sim21$.  Fainter still, is the Cal\'{a}n-ESO
proper motion catalog \citep{ruiz01}, which contains $542$ stars with
$\mu>200\,\masyr$ to $R\sim 19.5$ in a $350$ square degree field in the
south.  Also in the south is the \citet{wroblewski01} survey of 147
stars with $\mu>150\,\masyr$ ($\smu \sim 6-22\,\masyr$) over a 25 square
degree field down to $B\leq 18.5$.  Using the Digitized Sky Survey
(DSS), \citet{lepine03} find proper motions for stars between $500\leq
\mu\leq 2000\,\masyr$ accurate to $10\,\masyr$ with $8\leq R \leq20$
covering over $98\%$ of the northern sky.

Finally, several proper motion catalogs have been produced as byproducts
of searches for halo dark matter and therefore have wholly
different selection functions than the surveys mentioned above.  The
EROS 2 HPM catalog \citep{eros99} selects stars from 413 square
degrees at high Galactic latitude to $V\leq21.5$ and
$I\leq20.5$ with an accuracy of $25\,\masyr$.  The MACHO survey
provided a catalog of 154 high proper-motion ($\mu > 5 \,\masyr$) stars
from a total of 50 square degrees in fields towards the Galactic bulge
and the LMC with an accuracy of $\sim 0.8 \,\masyr$.  The Optical
Gravitational Lensing Experiment (OGLE II) catalog \citep{sumi03}
yielded proper motions for 5,078,188 stars with $\mu\leq 500 \,\masyr$
and accuracy $\smu \sim 0.8-3.5\,\masyr$ from a total imaging area of 11
square degrees towards the Galactic bulge.

These catalogs have fueled many different veins of study.  Searches
for a variety of distinct stellar populations including nearby stars
\citep{reid01, jahr01, scholz01, gizreid97, henry97}, subdwarfs
\citep{gizreid97, ryan92, digby03}, white dwarfs \citep{reid01b,
schmidt99, liebert79, jones72, luyten7077}, brown and L dwarfs
\citep{eros99}, halo stars \citep{gould03d}, and wide binaries
\citep{chaname03} are all most effectively carried out using proper
motion catalogs such as those described above.  Proper motion catalogs
have also been useful as a way to select candidates for future
microlensing events detectable with next generation surveys 
\citep{salim00} such as the {\it Space Interferometry Mission}
as well as candidates for planetary transit searches
\citep{gould03b}.  Searches for dark matter in the halo in
the form of white dwarfs \citep{opp01,reid01b} and MACHOs
\citep{yoo03} have also benefited from proper motion catalogs.
Finally, the structure of the stellar halo including the velocity
ellipsoid parameters \citep{gould03d} and the granularity of the
stellar halo \citep{gould03c} have both been made possible by the
large samples of halo stars culled from high proper-motion catalogs.

In this paper, we construct a new proper-motion catalog by combining
data from the Data Release One (DR 1) of the Sloan Digital Sky Survey
(SDSS, http://www.sdss.org) and USNO-B1.0 \citep{monet03}.  
SDSS contains photometry and
proper motions for stars over several disjoint areas at high
Galactic latitude totaling several thousand square degrees.  SDSS assigns
proper motions to most of these stars by identifying each with the nearest
entry in USNO-A2.0 \citep{usnoa2}, a position catalog based on
circa 1950 simultaneous blue and red photographic plates.  Many of these
proper motions are spurious, either because the SDSS star has
no counterpart in USNO-A (and so is identified with an unrelated star)
or because it has moved so far that the nearest USNO-A entry is not
the true counterpart of this star.  Moreover, these proper motions
suffer systematic errors due to small but significant zonal biases
in USNO-A.  USNO-B also contains many spurious high proper-motion
entries that arise from inadvertent association of unrelated stars
that have moved significantly between plate epochs, and it also
suffers from zonal errors.

Here we use the SDSS quasars to remove the zonal errors from
each of these two source catalogs.  Because the SDSS spectroscopic
survey covers only about half the area of the photometric survey,
our catalog is restricted to this smaller region in which the
quasars are reliably identified.  We then combine the two corrected
proper-motion measurements to form the single best estimates of the
proper motions, which have errors of $\sigma_\mu\sim 4\,\masyr$.

The very process of cross-identifying SDSS and USNO-B eliminates
most of the spurious entries in each.  We develop a graphical
technique to estimate the remaining contamination in various
subsets of the catalog in order to assist in the formulation
of selection procedures appropriate to various applications.
Our final catalog contains 345,000 entries with 
proper motions $\mu\geq 20\,\masyr$ and magnitudes $r'\leq 20$.

\citet{digby03} were the first to carry out a cross-identification of
SDSS with an independent proper-motion catalog, namely SuperCOSMOS.
They did not attempt to combine the two proper-motion measurements,
but rather used the SDSS proper motions as a cross check.  They
were mostly interested in combining SDSS photometry with their
proper motions in order to produce a reduced proper motion diagram
from which halo stars could be identified.  They achieved a
precision of $\sigma_\mu\sim 8\,\masyr$ and set their catalog threshold at
$\mu\geq 40.5\,\masyr$.

	In \S~2 we describe our initial SDSS sample selection.
In \S~3 we develop our method for correcting SDSS and USNO-B
proper motions using SDSS quasars, and we use these quasars
to evaluate the error properties of both these catalogs as well
as of our final adopted proper motions.  In \S~4 we describe the
construction of the stellar proper-motion catalog, and in \S~5
we develop visual techniques for understanding its statistical
properties.  We also present reduced proper motion diagrams.
Finally, in \S~6 we describe the electronic catalog.

\section{SDSS Selection
\label{sec:sdsssel}}

We select stars and quasars from the SDSS DR1
using the online query tools provided by SDSS.  
(For technical references, see \hfil\break\noindent
http://www.astro.psu.edu/users/dps/sdsstechrefs.html.)
We select all quasars that have matches within $2\arcsec$ to
corresponding sources within the USNO-A catalog and that are
spectroscopically confirmed quasars . For our star sample, we 
initially select all objects within DR1 that 1) are classified as 
stars by SDSS photometrically, 2) have proper motions 
(measured from USNO-A and SDSS) in the range $7\,\masyr \leq \mu \leq 
420\,\masyr$, and 3) have magnitudes $r'< 22$.    We then supplement
this with an additional query covering $\mu>420\,\masyr$ but restricted
to $r'< 20$.

\section{Quasar-Based Calibration
\label{sec:qso}}

We use the quasars for three related purposes.  First, to remove
the (position-dependent) astrometric biases in SDSS/USNO-A and USNO-B
proper motions.  Second, to measure the errors in these proper motions
and the covariances among them.  Third, as a benchmark of accurately
identified objects against which we can compare the distributions
of variously selected stellar subsamples in order to judge whether
they are significantly contaminated by spurious objects.  The
quasars have two characteristics that permit them to perform these
three functions.  First, they are spectroscopically confirmed by SDSS.
Second, they have known proper motions: namely zero.

\subsection{Individual Catalog Analyses
\label{sec:separate}}

We begin by correcting the SDSS proper motions by using neighboring
quasars to measure the local proper-motion bias.  
We determine this bias by taking the
weighted average of all other quasars,
\begin{equation}
\vec\mu^m_\bias = {1\over w^m_\tot}\sum_{n\not=m}w^m_n\vec\mu^n,\quad
w^m_\tot = \sum_{n\not=m} w^m_n,\quad
w^m_n \equiv \exp[-(\vec\theta^m - \vec\theta^n)^2/2\theta_0^2],
\label{eqn:bias}
\end{equation}
where $\vec\theta^n$ is the vector position of the $n$th quasar and
$\theta_0$ is a smoothing length.  We find that putting 
$\theta_0=0.75\,$deg minimizes the residual scatter.  In addition
to excluding the star that is being corrected from the sum, we also
exclude all those with $R_\usno>19.5$ or $R_\usno<15$, 
although this has only a very
minor effect.  We find that this correction improves the scatter
in the quasars by
\begin{equation}
\label{eqn:scattersdss}
(\sigma_\alpha,\sigma_\delta)_\s:\quad
(5.2,5.8)\,\masyr\rightarrow(4.7,4.7)\,\masyr\qquad \rm (SDSS),
\end{equation}
for the sample of 14,740 quasars.

We then apply the same procedure to USNO-B.  Of course, we exclude
all quasars that are not detected in USNO-B, but we also exclude
those that are recorded as having proper-motion errors exceeding 
$10\,\masyr$ in either direction, as well as those
with zero errors in both components.  About 99\% of these latter
objects are listed as having exactly zero proper motion.
Since, as we show below, the bias at the location of these quasars
is decidedly non-zero, these entries must be catalog artifacts rather
than highly precise measurements.  These two cuts reduce the sample
to 3543 quasars, with more than 90\% of the reduction being the
exclusion of the zero-error quasars.  Finally, because the quasar
sample is now fairly sparse, it is possible that some quasars do
not have enough near neighbors to accurately measure their bias.
We therefore exclude an additional 175 quasars with $w^m_\tot <3$,
leaving a sample of 3368.  For these,
\begin{equation}
\label{eqn:scatterusno}
(\sigma_\alpha,\sigma_\delta)_\u:\quad
(7.9,7.1)\,\masyr\rightarrow(5.1,5.0)\,\masyr\qquad \rm (USNO-B).
\end{equation}
Thus, the local bias (and hence the improvement when this bias
is removed) is substantially larger for USNO-B than for SDSS/USNO-A.
When we measure the bias for the zero-error quasars, it is similar,
thus motivating their above-stated exclusion from the analysis.

\subsection{Joint Catalog Analysis
\label{sec:joint}}

Since we will eventually measure proper motions for our stellar sample
using both SDSS/USNO-A and USNO-B, it is important to understand the
joint error properties of all four quantities being measured, which
we group together as a vector $a_i$,
\begin{equation}
a_i^m \equiv (
\mu_{\alpha,\cor,\s}^m,
\mu_{\alpha,\cor,\u}^m,
\mu_{\delta,\cor,\s}^m,
\mu_{\delta,\cor,\u}^m),
\label{eqn:adef}
\end{equation}
where 
$\mu_{\alpha,\cor,\s}^m\equiv \mu_{\alpha,\s}^m -\mu_{\alpha,\bias,\s}^m$,
and similarly for the $\delta$ component.
We first evaluate the covariances $c_{ij}$ of the $a_i$ averaging over
the 3368 quasars common to both samples,
\begin{equation}
c_{ij} \equiv \l a_i a_j\r= {1\over 3368}\sum_{m=1}^{3368} a_i^m a_j^m.
\label{eqn:cijdef}
\end{equation}
Note that while one would usually write  
$c_{ij}=\l a_i a_j\r -\l a_i\r\l a_j\r$, in the
present case the $\l a_i\r$ are known a priori to vanish.
We find errors
\begin{equation}
\sqrt{c_{ii}} = (4.13,5.15,4.27,4.96)\,\masyr, 
\label{eqn:cii}
\end{equation}
and correlation coefficients,
\begin{equation}
{c_{ij}\over\sqrt{c_{ii}c_{jj}}}=
\left(\matrix{1.000 & 0.551 & -0.034 & -0.052\cr
              0.551 & 1.000 & -0.065 & -0.080\cr
             -0.033 & -0.065 & 1.000 & 0.554\cr
             -0.052 & -0.080 & 0.554 & 1.000\cr}
\right).
\label{eqn:cij}
\end{equation}
Note that the SDSS scatter is slightly reduced compared to 
equation (\ref{eqn:scattersdss}), indicating that the
quasars accessible to USNO-B are somewhat better behaved than
the full sample.  Since we will also only be considering stars
with entries in both catalogs, the numbers quoted in
equation (\ref{eqn:cii}) are the directly relevant ones.

Equation~(\ref{eqn:cij}) shows that the SDSS/USNO-A and USNO-B
measurements are correlated at the 55\% level.  Presumably, this
is due to the fact that the first Palomar Observatory Sky Survey
(POSS I) underlies the first (circa 1950) epoch of both measurements.
Also note that the correlations between the $\alpha$ and $\delta$
components, while small, are statistically significant because the
uncertainties in these quantities are $(3368)^{-1/2}=0.017$.  We therefore
keep these cross terms in our calculations, which we describe in the 
appendix.  However, while this mathematical analysis is straightforward,
it is a bit dense.  We therefore present
a simplified treatment here that ignores these small terms and so is
much more transparent while yielding almost exactly the same results

  The internal error
is estimated by determining how well the SDSS/USNO-A and USNO-B
measurements agree with each other, while the external error
is estimated by determining how well the combined proper-motion 
measurements agree with reality (i.e., that the proper motions are 
actually zero).  

The internal error is characterized by,
\begin{equation}
\chi^2_{\inter,m} = 
\biggl[{\Delta \mu^m_{\alpha}\over \sigma(\Delta \mu_{\alpha})}\biggr]^2
+\biggl[{\Delta \mu^m_{\delta}\over \sigma(\Delta \mu_{\delta})}\biggr]^2,
\label{eqn:chi2intdefQ}
\end{equation}
where $\Delta\mu^m_{\alpha}=a_1^m - a_2^m = 
\mu_{\alpha,\cor,\s}^m-\mu_{\alpha,\cor,\u}^m$, where
$\sigma(\Delta \mu_{\alpha})$ is the uncertainty in this quantity, and
similarly for the $\delta$ component.  Since the two measurements
are correlated, the error in their difference is given by
$[\sigma(\Delta \mu_{\alpha})]^2 = c_{11}-2 c_{12}+c_{22}$.  Substituting
the values given in equations~(\ref{eqn:cii}) and (\ref{eqn:cij})
yields
\begin{equation}
[\sigma(\Delta \mu_{\alpha}),\sigma(\Delta \mu_{\delta})]
=(4.49,4.40)\,\masyr,
\label{eqn:CklevalQ}
\end{equation}
That is, because the correlation coefficients between the SDSS/USNO-A and
USNO-B measurements (eq.~[\ref{eqn:cij}]) are so large, the errors in 
the differences of these measurements are only of order the errors in
each measurement separately (rather than being multiplied by $2^{1/2}$).

\begin{figure}
\plotone{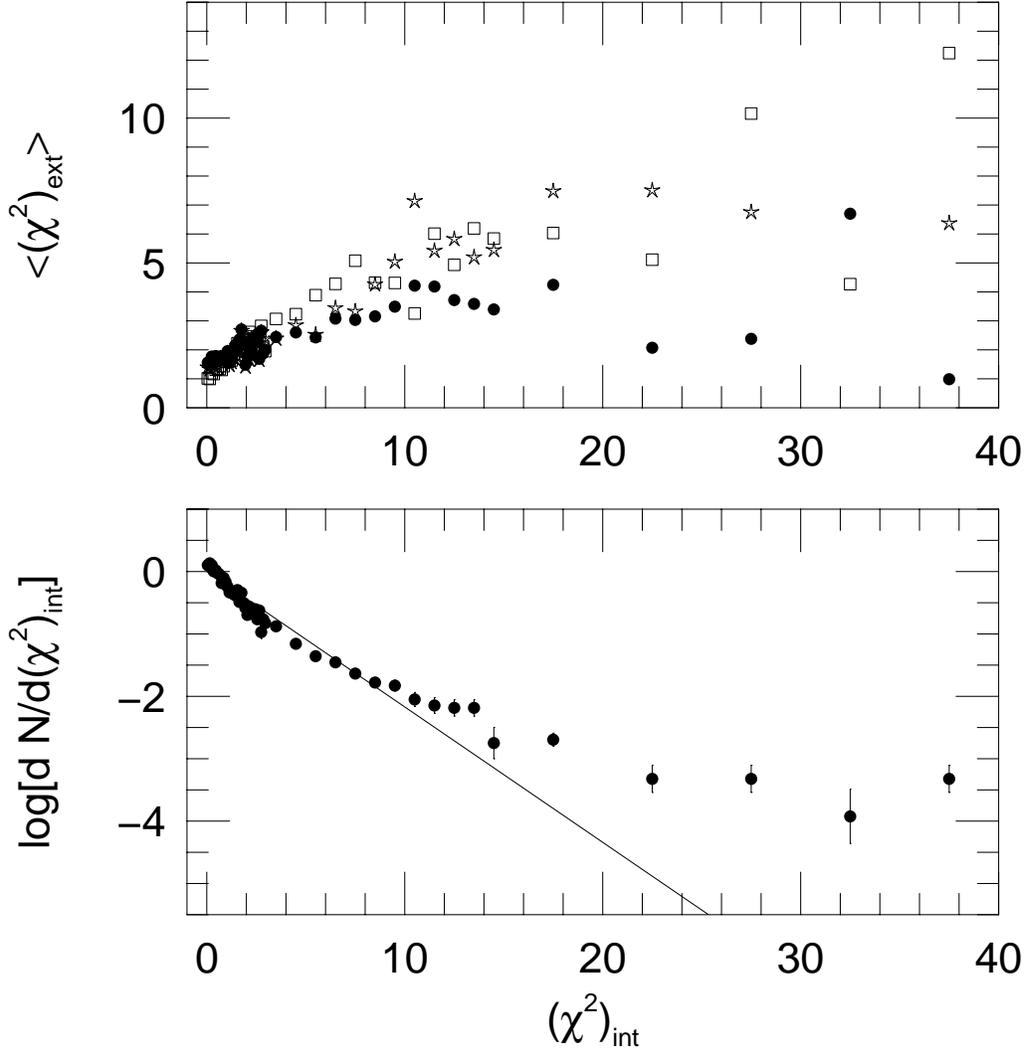}
\caption{\label{fig:chi2}
Lower panel:  Distribution of the internal-error indicator $\chi^2_\inter$ 
(derived from the difference between SDSS and USNO-B proper-motion
measurements, see eq.~[\ref{eqn:chi2intdef}]) for $N=3368$ quasars.  The
distribution is normalized by $N/2$ to permit easy comparison with
other distributions (see Fig.~\ref{fig:chi2chart}).  
The straight line is the expected distribution
assuming Gaussian statistics.  It is not a fit: it has no free parameters.
The distribution is Gaussian in its core ($\chi^2_\inter\la 8$) but
deteriorates significantly in its wings.
Upper panel:  Mean external-error indicator $\chi^2_\ext$ 
(derived from the differences between the measured proper motions and their
known zero value, see eq.~[\ref{eqn:chi2ext}]) 
as a function of $\chi^2_\inter$ for the same sample.  Values for
the SDSS-only ({\it stars}), USNO-B-only ({\it open squares}) and
combined ({\it filled circles}) measurements are shown separately.
In the inner Gaussian core  $\chi^2_\inter\la 4$, all three cluster
around their expected values of 1, 1, and 2 respectively.  At higher
$\chi^2_\inter$, for which SDSS and USNO-B are in increasing
conflict and therefore cannot both be correct, $\chi^2_\ext$ rises
as would be expected.  Surprising, however, for the combined measurements,
the rise saturates at $\chi^2_\ext\sim 4$, perhaps indicating that
large-error measurements of SDSS and USNO-B are anti-correlated.
}\end{figure}

The distribution of $\chi^2_\inter$ for the 3368 quasars is shown in the 
lower panel of Figure~\ref{fig:chi2}.  The straight line is the prediction for
Gaussian statistics, i.e., a $\chi^2$ of two degrees of freedom,
which is an exponential distribution, $f(\chi^2)=\exp(-\chi^2/2)$.
This line is not a fit: it has no free parameters.  The figure shows
that the internal errors are Gaussian distributed in the core,
$\chi^2_\inter \la 8$, and then deteriorate to a much flatter distribution
in the wings, $f(\chi^2)\propto \exp(-\chi^2/7)$.  This deterioration
is not unexpected and is typical of many types of astronomical observations.

Next we evaluate the external errors.  Let $\tilde \mu_\alpha^m$ be the
best estimate of $\mu^m_\alpha$ given the two correlated measurements
from SDSS and USNO-B, respectively.  Let $\sigma(\tilde \mu_\alpha)$
be the uncertainty in this quantity, and similarly for the $\delta$ component.
Since the quasar proper motion is known to vanish, these ``best estimates''
are exactly equal to the error in the measurement.  Hence, the
external errors are characterized by,
\begin{equation}
\chi^2_{\ext,m} = 
\biggl[{\tilde \mu^m_{\alpha}\over \sigma(\tilde \mu_{\alpha})}\biggr]^2
+\biggl[{\tilde \mu^m_{\delta}\over \sigma(\tilde \mu_{\delta})}\biggr]^2.
\label{eqn:chi2extQ}
\end{equation}
For two correlated measurements $a_1$ and $a_2$ of the same quantity, the
best combined estimate and its uncertainty are given by,
\begin{equation}
\tilde a = 
{a_1(c_{22} - c_{12}) + a_2(c_{11} - c_{12})\over c_{11} - 2c_{12} + c_{22}}
\qquad
\tilde\sigma^2 = {c_{11}c_{22} - c_{12}^2\over c_{11} - 2c_{12} + c_{22}}.
\label{eqn:tildeaQ}
\end{equation}
These equations can be derived using the formalism presented in the Appendix.
However, it is easy enough to verify that they reduce to familiar results
when {\it either} $c_{12}=0$ {\it or} $c_{11} = c_{22}$.  Substituting the
values in equations~(\ref{eqn:cii}) and (\ref{eqn:cij}), yields
\begin{equation}
[\sigma(\tilde \mu_{\alpha}),\sigma(\tilde \mu_{\delta})]
=(3.96,4.00)\,\masyr.
\label{eqn:tildecijevalQ}
\end{equation}
So, just as the error in the difference of measurements is not
substantially bigger than the separate errors, so the error in
the combined measurement is not substantially smaller.  Again this is because
of the strong correlation between the SDSS/USNO-A and USNO-B
proper-motion measurements.

The upper panel of 
Figure~\ref{fig:chi2} shows the mean $\chi^2_\ext$ as a function
of $\chi^2_\inter$.  In the inner core ($\chi^2_\inter<4$), which
accounts for about 70\% of the sample, $\l\chi^2_\ext\r\sim 2$,
i.e., its expected value.  It grows beyond that, as one would expect,
because if $\chi^2_\inter$ is high, then at least one of the measurements
must be seriously in error, which should then corrupt the combined
estimate.  However, the rise unexpectedly seems to saturate
at $\l\chi^2_\ext\r\sim 4$, so that the most discrepant measurements in 
SDSS/USNO-A
must be significantly anti-correlated with the the most discrepant ones in
USNO-B.  Further insight is gained by considering the behavior
of the external errors of each catalog measurement separately
(still plotted as a function of the combined internal error).
These show similar behavior, indicating that they are about 
equally responsible for the problem measurements.  They also
track the combined external error in the Gaussian part of the
distribution ($\chi^2_\inter \la 8$), but at higher $\chi^2_\inter$
the individual external errors continue to rise roughly linearly,
while the external errors flatten.  The deviation is pronounced
only for $\chi^2_\inter \ga 16$, which encompasses only about
35 quasars.  It will be interesting to see if this pattern is
maintained when larger areas of the SDSS catalog are made
available.

\section{Stellar Proper Motion Catalog
\label{sec:stellarpm}}

\subsection{Construction
\label{sec:construct}}

It is now straightforward to construct the catalog.
We begin with our full sample of $5\times 10^6$ stars selected from
the SDSS catalog that satisfy $\mu \geq 7\,\masyr$ and $r'<22$,
but excluding faint ($r'>20$), high proper-motion ($\mu>420\,\masyr$)
stars for reasons described below.
We first correct these proper motions for bias in the same way that we
did for the quasars.  See equation~(\ref{eqn:bias}).  However, since
the quasars require spectroscopic confirmation, and the
SDSS spectrographic catalog covers a smaller section of the sky than
the photometric catalog, there are many stars that are too far from
any quasars to be reliably de-biased.  Therefore, we accept stars only
if $w^m_\tot>4$.  At this point, having obtained a better estimate
of the proper motion, we further restrict the proper motion criterion
to $\mu>10\,\masyr$.  These two selection procedures, primarily the first,
reduce our sample by more than half to $1.9\times 10^6$.

We then search for the remaining stars in USNO-B, using a $2''$ search
radius to allow for both catalog errors and the motion of the star between
the SDSS epoch and the (2000) epoch of USNO-B.  We increase this radius
to $2.\hskip-2pt''5$ for stars with $\mu>420\,\masyr$.
As we did for the
quasars, we exclude all entries with zero errors in both components, 
with errors greater than $10\,\masyr$ in either component, or with
$w^m_\tot <3$.  This reduces the catalog to just under $9\times 10^5$.

Finally, we impose two additional cuts
\begin{equation}
r'\leq 20,\qquad \mu > 20\,\masyr 
\label{eqn:finalcuts}
\end{equation}
As we describe in \S~\ref{sec:tests}, the great majority of detections
with $r>20$ are spurious.  This is the reason for the first cut.
The proper motion threshold is set by demanding ``$5\,\sigma$''
detections.  If the statistics were Gaussian, this would imply
a probability $\exp(-25/2)\sim 10^{-5.4}$ that a zero-proper-motion
star could scatter into our sample.  However, as shown by 
Figure~\ref{fig:chi2}, the tail of the error distribution is decidedly 
non-Gaussian, with roughly 1\% of the quasars showing internal errors,
$\chi^2_\inter>16$.  While the initial evidence is that this large
tail of internal errors is not reflected in a similar tail of external
errors (see \S~\ref{sec:joint} and Fig.~\ref{fig:chi2}), we nevertheless
proceed conservatively and set a relatively high threshold to minimize
contamination.

\begin{figure}
\plotone{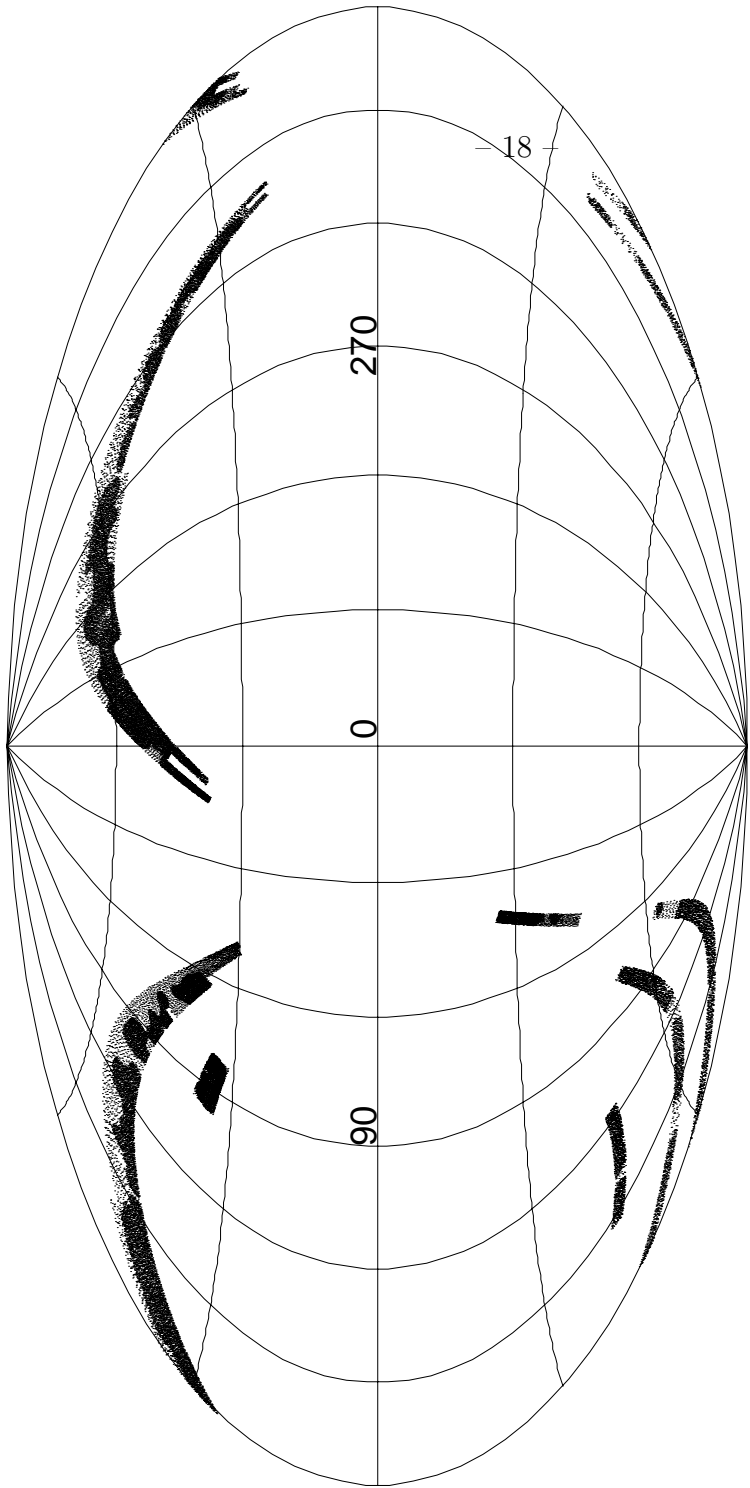}
\caption{\label{fig:aitoff}
Distribution of catalog stars on the sky.  The lightly shaded regions
are the original selection from the SDSS photometric database.  The
heavily shaded regions are the final sample.  Their smaller area
reflects the fact that the proper motions must be corrected using
SDSS quasars in the spectroscopic database, which has substantially
smaller sky coverage.
}\end{figure}

The sky coverage of the catalog is shown in Figure~\ref{fig:aitoff}.
The lightly shaded areas indicate the coverage of the photometrically
selected stars, while the darker areas are the regions of the catalog.
The difference is mostly due to the smaller region covered by the
spectroscopic catalog, but is also affected by the failure to recover
most quasars in USNO-B.

\section{Catalog Characterization
\label{sec:tests}}

\begin{figure}
\plotone{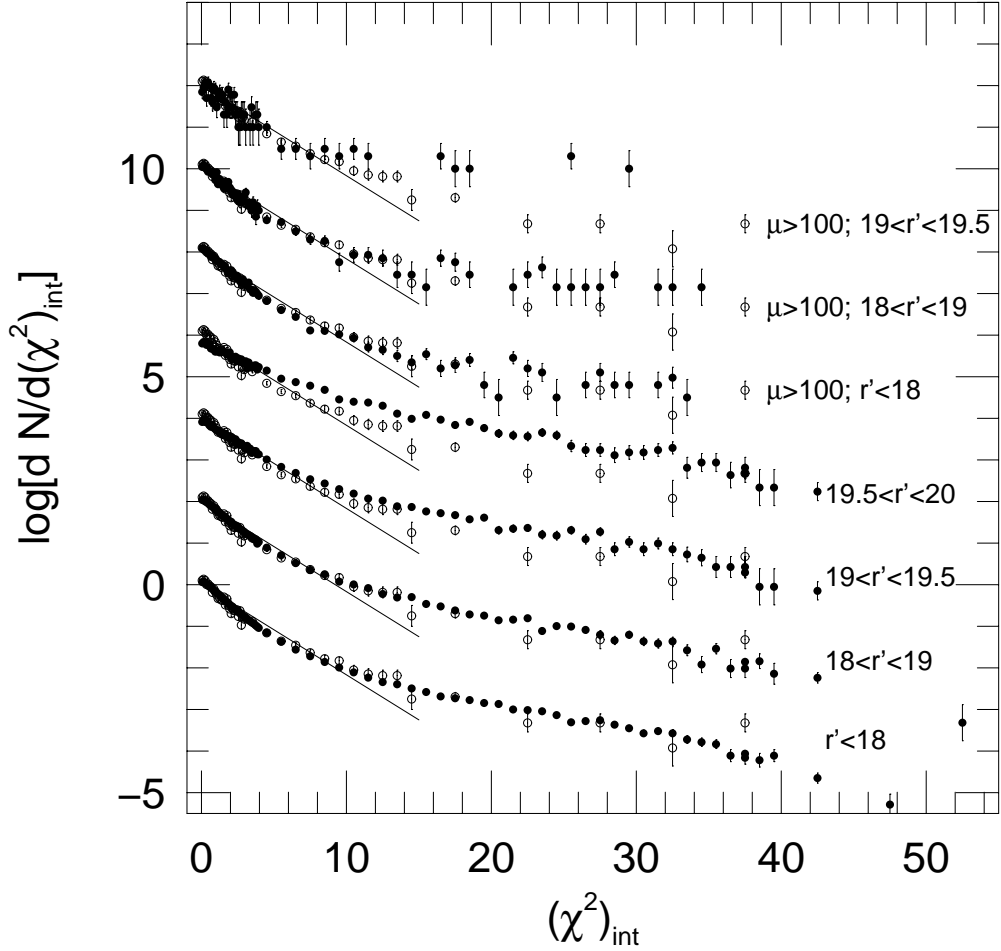}
\caption{\label{fig:chi2chart}
Distribution of internal-error indicator $\chi^2_\inter$ for various subsets
of the stellar catalog ({\it solid circles}) compared to that for
quasars ({\it open circles}, see Fig.~\ref{fig:chi2}).  The three upper
tracks are restricted to $\mu>100\,\masyr$, while the four
lower tracks go down to the catalog limit of $\mu=20\,\masyr$.
Note that the $r'<18$ distributions track the quasars extremely
well, indicating that their error distributions are similar and
that there are few spurious objects.  However, the faintest bin,
$19.5<r'<20$, has a tail that is an order of magnitude higher
than the quasars, probably indicating a large number of spurious
objects, and this effect is even more pronounced among the high
proper-motion stars.
}\end{figure}

In order to gain insight into the properties of the catalog as
a function of magnitude and proper motion, we compare in
Figure~\ref{fig:chi2chart} the internal error distribution
of the quasars ({\it open circles}) 
to that of various subsets of the stars ({\it filled circles}).  For
example, the lowest track compares the quasars to the stars with
$\mu\geq 20\,\masyr$ and $r'<18$.  The two distributions are essentially
the same, which corroborates one's expectation 
that the errors are the same and that there
are not a large number of spurious entries in this subsample.  This
is to be compared with the middle track, which shows
$\mu\geq 20\,\masyr$ and $19.5<r'<20$.  This sample has about an order of
magnitude increase in the fraction of stars with large internal errors,
$\chi^2_\inter \ga 20$.  Hence, either there are a large number of
spurious entries in this subset or the faint magnitudes induce unusually
large proper-motion errors in one or both catalogs.  Comparison of the
first and fifth tracks shows that the problem cannot be entirely
faint magnitudes.  These tracks represent the same magnitude range
$19<r<19.5$, but very different proper-motion cuts, $\mu>20\,\masyr$
and $\mu>100\,\masyr$, respectively.  The higher proper-motion
sample has a much higher rate of large internal errors.  Thus, it
is likely that spurious detections play a major role at faint magnitudes,
and more so when the proper motion is also high.

It is well-known that both SDSS and USNO-B separately contain large
numbers of spurious high proper-motion stars.  For USNO-B,
roughly 99\% of the entries with $\mu>180\,\masyr$ are spurious
\citep{gould03e}.  Since SDSS is merely matching to the nearest
USNO-A entry, spurious entries are bound to occur, especially when
the star is too faint to be recovered by USNO-A.  While the chance
that spurious entries will coincide in both catalogs within our $2''$
search radius is small in each individual case, there are literally
of order $10^6$ trials.

The implication is that one must carefully select from the catalog
in accordance with one's goals and guided by Figure~\ref{fig:chi2chart}.
If for example, one wants a very clean sample, one might accept only
stars in each magnitude range up to the maximum $\chi^2_\inter$ to
which the stars track the quasars.  On the other hand, if one were
looking for common proper motion companions to brighter
stars, one might search up to $r'\leq 20$, recognizing that additional
confirmation would be required for any candidates thus detected.

\begin{figure}
\plotone{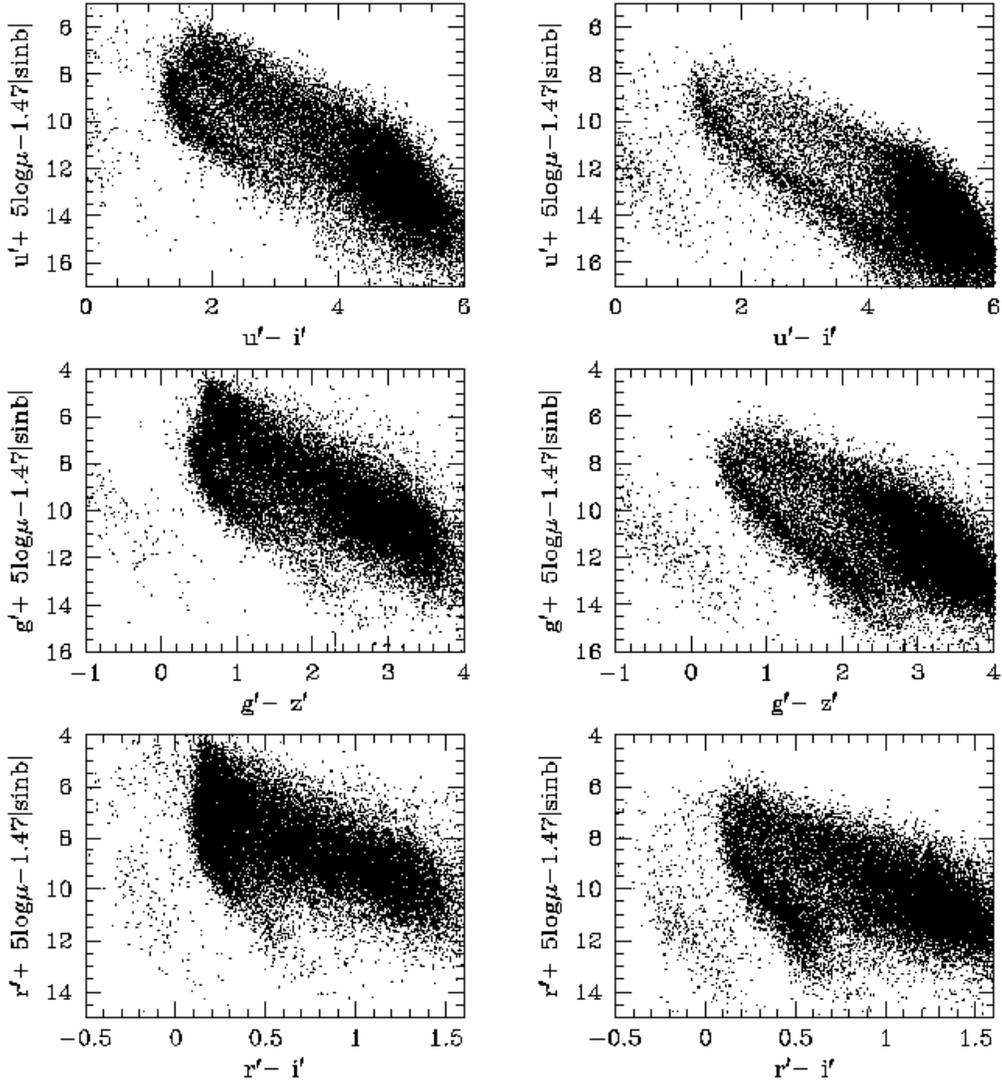}
\caption{\label{fig:rpm}
Reduced proper-motion diagrams for catalog stars $r'<19$ for 
stars down to the proper-motion limit of $20\,\masyr$ (left panels)
and restricted to relatively high proper motions, $\mu\geq 5\,\masyr$
(right panels).  All panels have about 25,000 points, meaning that
the right panels are drawn from a larger fraction of the catalog.
The right panels are relatively enhanced in white dwarfs (lower left),
red dwarfs (lower right), and subdwarfs (lower diagonal track).
Note also how the subdwarf morphology evolves with color.
}\end{figure}

Figure~\ref{fig:rpm} contains six reduced proper motion (RPM)
diagrams.  These have $\mu\geq 20\,\masyr$ and $\mu\geq 50\,\masyr$
thresholds for the left and right sides, respectively.  All
diagrams are restricted to $r'<19$.  The ordinates in these
diagrams are the adjusted RPMs, e.g.,
\begin{equation}
\eta_{r'} \equiv r' + 5\,\log{\mu\over 1''\, {\rm yr}^{-1}} -1.47|\sin b|,
\label{eqn:etadef}
\end{equation}
where $b$ is Galactic latitude.  The first two terms are the
standard RPM.  These are related to the absolute magnitude by,
\begin{equation}
H_{r'} = r' + 5\,\log{\mu\over 1''\, {\rm yr}^{-1}} = 
M_{r'} + {v_\perp\over 47.4\,\kms} + A_{r'},
\label{eqn:absmag}
\end{equation}
which causes an RPM diagram to simulate a color-magnitude diagram.
Here $v_\perp$ is the transverse velocity and $A_{r'}$ is the
extinction.  However, \citet{salim03} showed that while RPM diagrams 
at different $b$ look similar, they are vertically offset from one another, .
They therefore introduced $\eta$, which has a $b$-dependent term designed to
compensate for this effect.

The figure illustrates the changing morphology of RPM diagrams with
choice of color and proper-motion threshold.  In the ``high''
threshold sample, the white dwarfs (lower left), red dwarfs
(lower right), and subdwarfs (lower diagonal track) are all more
pronounced.  We have placed ``high'' in quotation marks because this
sample has a lower proper-motion threshold than any previous RPM
diagram at these magnitudes, except that of \citet{digby03} to
which it is comparable.  See in particular their figures 6 and 7.
The low-threshold sample has a higher density of upper-main-sequence
and turnoff stars.  (The two sets of diagrams are drawn respectively
from 9\% and 60\% of the catalog, so that they have a similar number
of points, roughly 25,000.)\ \  
Note also that in $u'-i'$, the red subdwarfs tend
to merge with the main-sequence red dwarfs, while in $r'-i'$, the
red ends of these tracks remain distinct.

\section{Catalog Description
\label{sec:description}}

The first four columns of the catalog give the right ascension
and declination of the star as reported in SDSS and USNO-B,
respectively.  Both are equinox 2000.  USNO-B is also epoch 2000,
while SDSS is at the epoch of the observations (which are circa 2000).
Columns (5) and (6) give the east and north components of
$\vec\mu_{\cor,S}$, the corrected SDSS proper motion as described
by equations (\ref{eqn:bias}) and (\ref{eqn:adef}).
Columns (7) and (8) give $\vec\mu_{\cor,B}$, the corrected USNO-B
proper motion, while columns (9) and (10) give the weighted
average of the two as described by equation~(\ref{eqn:tildeai}).
Column (11) is the scalar proper motion, i.e., the root-sum-square
of columns (9) and (10).  Columns (12) through (16) give SDSS PSF
photometry in $u'$, $g'$, $r'$, $i'$, and $z'$, while columns (17)
and (18) give the blue and red magnitudes of the USNO-A star that
SDSS has associated with this entry.  Column (19) gives $H_{r'}$,
the unadjusted $r'$ band RPM (see eq.~[\ref{eqn:absmag}]), while
column (20) gives the adjusted RPM, $\eta_{r'}$ 
(see eq.~[\ref{eqn:etadef}]).  Columns (21) and (22) give
$w_{\tot,S}$ and $w_{\tot,B}$, the total statistical weights of the
quasars used in the local proper-motion correction.  
See equation~(\ref{eqn:bias}).  Finally, column (23) gives the
internal-error indicator $\l \chi^2_\inter\r$ as described by
equation~(\ref{eqn:chi2intdef}).

The fortran format statement for the catalog is\hfil\break\noindent
(4f10.5,7f7.1,5f7.3,2f5.1,2f7.3,2f5.1,f7.2).
The catalog can be found at\hfil\break\noindent
http://www-astronomy.mps.ohio-state.edu/$\sim$gould/SDSS-USNOB/cat.dat.gz

If the catalog is restricted to regions in which the stars track
the quasars in Figure~\ref{fig:chi2chart}, 
then the errors can be expected to be similar to
those of the quasars, namely $\sigma_\mu= 4.0\,\masyr$ in each
direction.


\acknowledgments 
We thank D. Monet and the USNO-B team for providing us with a copy of
the USNO-B1.0 catalog.  Support for the SDSS Archive comes from
the Alfred P. Sloan Foundation, the Participating Institutions, 
NASA, the NSF, the DoE, the Japanese Monbukagakusho, and the 
Max Planck Society.  AG was supported by grant AST 02-01266 from the NSF 
and by JPL contract 1226901.  JAK was supported by an OSU Dean's
Distinguished Fellowship.

\appendix

\section{Covariant forms of $\chi^2_\inter$ and $\chi^2_\ext$
\label{sec:appa}}

In \S~\ref{sec:joint}, we evaluated $\chi^2_\inter$ and $\chi^2_\ext$
under the simplifying assumption (almost realized in practice) that
the proper motion measurements in the $\alpha$ and $\delta$ directions
are not correlated.  Here we present a more general treatment.  
The reader is referred to \citet{gouldan} or \citet{gould03f} for the
mathematical underpinnings of this appendix.

We first introduce the two-vector $A^k$, defined by,
\begin{equation}
A^k \equiv \kappa^k_i a_i = (a_1-a_2,a_3-a_4),
\label{eqn:Adef}
\end{equation}
where the $2\times 4$ constraint matrix $\kappa^k_i$ is defined by
$\kappa^1_i=(1,-1,0,0)$ and $\kappa^2=(0,0,1,-1)$, and where
we have introduced the Einstein summation convention.  It is straightforward
to show that the covariances of the $A^k$ are given by
\begin{equation}
\cov(A^k,A^l) = C^{kl},\qquad C^{kl}\equiv \kappa^k_i c_{ij} \kappa^l_j.
\label{eqn:Ckl}
\end{equation}
Hence, the internal errors are characterized 
by,
\begin{equation}
\chi^2_{\inter,m} = A^l_m B^{kl} A^l_m,\qquad B\equiv C^{-1}.
\label{eqn:chi2intdef}
\end{equation}
Explicit evaluation of $C^{kl}$ using equations~(\ref{eqn:cii}) and
(\ref{eqn:cij}),
\begin{equation}
\sqrt{C^{kk}}=(4.49,4.40)\,\masyr,\qquad 
{C^{12}\over\sqrt{C^{11}C^{22}}} = -0.01.
\label{eqn:Ckleval}
\end{equation}
shows that it is almost exactly the same as given in the naive analysis
of \S~\ref{sec:joint}.

Next we evaluate the external errors.  The combined measurement of
the proper motion is found by imposing the constraint 
$\kappa^k_i a_i^m \rightarrow 0$, which yields best estimates for the
parameters, $\tilde a_i^m$,
\begin{equation}
\tilde a_i^m = a_i^m - D^k_m\alpha^k_i,\quad D^k_m \equiv B^{kl}A^l_m,
\quad \alpha^k_i \equiv c_{ij}\kappa^k_j
\label{eqn:tildeai}
\end{equation}
and covariances,
\begin{equation}
\tilde c_{ij} = c_{ij} - \alpha^k_i B^{kl}\alpha^l_j.
\label{eqn:tildecij}
\end{equation}
There are only two independent components of $\tilde a_i$ 
(one for each direction), and so $\tilde c_{ij}$ is effectively
a $2\times 2$ matrix.  We therefore compress their representations,
\begin{equation}
\tilde a_{2i-1} \rightarrow \tilde a_i,\qquad
\tilde c_{2i-1,2j-1} \rightarrow \tilde c_{ij},
\label{eqn:compress}
\end{equation}
and evaluate
\begin{equation}
\sqrt{\tilde c_{ii}}=(3.96,4.00)\,\masyr,\quad
{\tilde c_{12}\over\sqrt{\tilde c_{11} \tilde c_{22}}}= -0.063.
\label{eqn:tildecijeval}
\end{equation}
Again, this is almost identical to the result obtained in the
naive treatment.  The external error is then characterized by,
\begin{equation}
\chi^2_{\ext,m} = \sum_{i,j=1}^2 \tilde a^m_i \tilde b_{ij}\tilde a^m_j,
\qquad \tilde b \equiv \tilde c^{-1}.
\label{eqn:chi2ext}
\end{equation}

\clearpage

\end{document}